# Speech Technology Services for Oral History Research

Christoph Draxler[1], Henk van den Heuvel[2], Arjan van Hessen[3],
Pavel Ircing[4], Jan Lehečka[4]

[1]BAS / Ludwig Maximilian Universität, München, [2]Radboud University,
[3]University of Twente, [4]University of West Bohemia, Pilsen
draxler@phonetik.uni-muenchen.de, henk.vandenheuvel@ru.nl, a.j.vanhessen@utwente.nl,
{ircing,lehecka}@kky.zcu.cz

**Abstract**
Oral history is about oral sources of witnesses and commentors on historical events. Speech technology is an important instrument to process such recordings in order to obtain transcription and further enhancements to structure the oral account In this contribution we address the transcription portal and the webservices associated with speech processing at BAS, speech solutions developed at LINDAT, how to do it yourself with Whisper, remaining challenges, and future developments.

## 1. Introduction

Oral history testimonies rely first of all on the audio and/or video capture of the recorded material, typically in the form of interviews. Here, the first challenge is converting the audio signal into a readable text adequately reflecting the spoken word. Automatic Speech Recognition (ASR) has been employed since around three decades to obtain initial transcriptions of oral history interviews. The output text typically calls for extensive manual correction often equalling or exceeding (!) an effort equivalent to starting with manual transcription from scratch (Gref, 2022), especially if recordings are characterized by overlapping and/or dialectal speakers, background noises or mediocre recording quality.

The impressive performance of large AI-based speech models, especially their robustness to noise, the large range of supported languages, and the option to adapt them to additional languages with relatively little extra training, is these days greatly facilitating the generation of adequate transcripts in research areas such as oral history where spoken language is a major source of information.

However, depending on the researcher's needs there remain (other) challenges such as appropriate speaker attribution, output of more fine-grained speech events such as hesitation sounds (*uh*), stutters, word truncations, etc.

We have a longstanding track record in speech technological support for oral history research (see also Scagliola et al., 2020) in CLARIN ERIC[1] where we have initiated a resource family page for oral history corpora[2] and a transcription portal for oral history recordings[3] and speech processing facilities at LINDAT[4] (the Czech node of CLARIN ERIC). In this contribution we will address a number of speech technology tools and solutions in these contexts, remaining challenges and future developments.

More specifically, we will address the transcription portal and the webservices associated with speech processing at BAS (section 2), speech solutions and beyond developed at LINDAT (section 3), do it yourself with Whisper (section 4), remaining challenges (section 5) and future developments (section 6).

## 2. Webservices at BAS

The Bavarian Archive for Speech Signals (BAS) provides a large number of multilingual speech processing web services for academic users:

https://clarin.phonetik.uni-muenchen.de/BASWebServices/interface

The services support 40+ languages, plus a language-independent mode based on phonemic transcripts. The list of available languages is accessible via drop down menus on the web page.

The following services may be of particular interest to Oral History scholars. They can be used without authentication.

- ChannelSeparator separates the individual channels of a stereo recording and retains in each channel only the voice of the dominant speaker.
- G2P (Grapheme to phoneme) converts an orthographic text to its phonemic representation. The service allows a customized specification of pronunciation rules for vernacular language, dialects, and common coarticulation phenomena (e. g. 'haben wir' (*we have*) → /hamva/ or /hama/ in German). These rules improve the performance of automatic word alignment.
- MAUS (Munich automatic segmentation) aligns an orthographic transcript in one of the available languages and regional variants) with the audio signal. The performance of MAUS depends on

---

[1] https://www.clarin.eu/content/clarin-nutshell
[2] https://www.clarin.eu/resource-families/oral-history-corpora
[3] https://speechandtech.eu/transcription-portal
[4] See https://lindat.cz/en/services

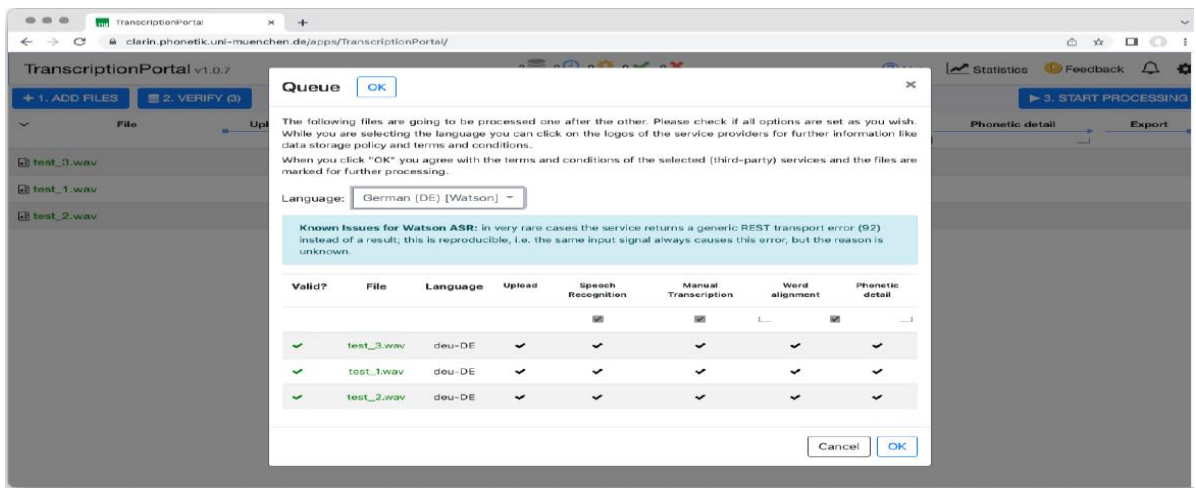

Figure 1: Selection of ASR language and processing steps in the transcription portal

- the language, the audio quality and the number of speakers; for German monologue recordings, it reaches 95% of the performance of human transcribers (Kipp et al. 1997).
- Octra is a graphical editor for orthographic transcription (Draxler & Pömp, 2022). It provides different views of the signal and the associated transcript, supports many input and output formats, and may split a recording into fragments to focus on relevant parts and/or distribute the workload. See further below for more details.
- Anonymizer automatically replaces the signal fragment corresponding to a given text by a beep, so that private information is effectively removed from both the recording and the transcript.

Automatic speech recognition is another web service. In contrast to the other services, it requires authentication as a member of academia, e.g. via the credentials necessary to login to a recognized academic institution (as recognized by CLARIN). The actual speech recognition is performed by external third-party providers, both academic and commercial, and thus may not be available if the privacy guidelines for a given recording or project do not allow data exchange with such providers.

Each service comes with a number of obligatory and optional parameters, e.g. the supported languages or file types, or output encodings and alphabets. and each service requires up- and downloading of data.

To reduce the number of file uploads and downloads, preconfigured service pipelines are available. This enables non-technical users to perform complex speech processing tasks without having to worry about low-level details. For example, the pipeline G2P→Chunker→MAUS→Anonymizer takes as input pairs of long audio files with an orthographic transcript plus a stop list of words, and returns a time-aligned transcript with the words from the stop list masked by a beep in the signal and a special symbol in the transcript. This pipeline works in any of the languages supported by G2P or MAUS.

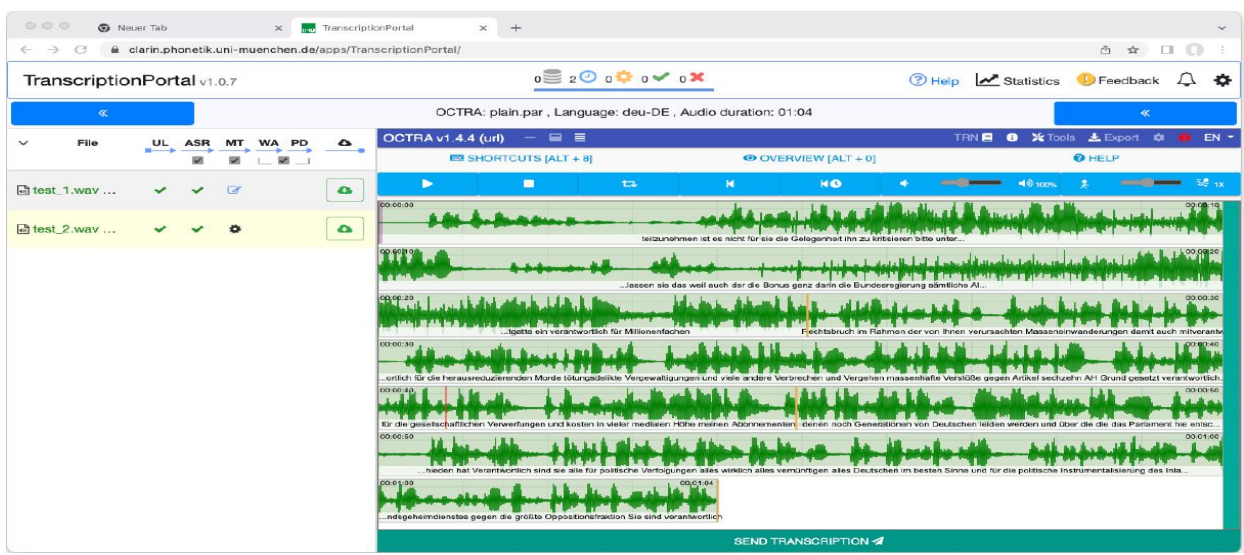

Figure 2: Multi-line preview of audio file and transcript in the Octra editor within the Transcription Portal

If the user has specified an email address, a link to download the results will be sent once processing is done – ideal for overnight processing.

Finally, all services feature help pages, and they may be used via REST calls from scripts and application programs. In fact, transcription editors such as ELAN (Wittenburg et al. 2006), EXMARaLDA (Schmidt & Wörner, 2014), or Octra access BAS web services in the background.

*Transcription portal*

The transcription portal was designed as a zero-configuration service for transcribing oral history recordings. Audio files are entered into the portal via drag & drop on the graphical user interface, The user then selects the language and which processing steps to apply (automatic speech recognition, manual correction, word alignment, export). See Figure 1.

The portal displays the status of every file in the workflow, and automatically calls all necessary services and tools in the workflow. See Figure 2.

At each step, the current state of the file can be examined and downloaded.

For subsequent in-depth analysis, the transcription portal supports exporting the transcript in a number of common formats, e.g. ELAN eaf, Praat TextGrid, or tabular or plain text for statistical and linguistic analysis.

Currently, the transcription portal relies on external service providers for the automatic speech recognition. In the ATRIUM project, AI-based speech recognition will be integrated into the portal. This will not only eliminate the need to access third party providers, but also increase the number of languages that can be accessed.

*Octra Backend*

Oral History recordings generally contain private information and thus strict requirements on data protection must be met. Octra Backend is a software for the management of transcription projects. It was designed with privacy as one of its key features. Octra backend operates in three scenarios:

a) in closed local area networks
b) in the intranet of a workgroup, a company or an institution, or
c) globally via the internet.

In all scenarios, only registered users may access data, and access is regulated by roles in a project. In scenario a), access is restricted to known machines in the local network, while in scenarios b) and c) privacy is ensured by encrypted communication. Audio and transcription data is managed by Octra Backend in its own protected file space, with file names hidden from outside viewers.

In Octra Backend, project administrators define tasks, e.g. manual correction of transcripts generated by ASR, or creating transcripts from scratch, and can assign these tasks to specific transcribers. Transcriptions are performed via Octra in the browser, and there is no need to open or save files, increasing process efficiency and reducing error-prone manual interactions.

The ATRIUM-project[5] starting in 2024 is an EU-funded infrastructure project targeted at archaeologists, with a section devoted to processing spoken language, namely for transcribing Oral History interviews. This task will focus on improving the user experience and the transcription performance, both in terms of quality and efficiency, of the existing transcription portal developed and maintained by BAS.

## 3. LINDAT for Oral Historians

The web-based ASR engine named UWebASR[6] has been deployed as a service within the LINDAT/CLARIAH-CZ portal in 2018. In order to be able to access this service – and other services that are available in the portal – the user has to authenticate herself/himself as a member of academia (in the same manner as for the use of BAS services mentioned above). The aim is to always provide the best possible ASR performance – we have therefore switched the underlying technology to state-of-the-art wav2vec models (Baevski, 2020) recently. The service is provided for recordings in English, Czech, Slovak and German. All the language-specific models are built from pre-trained models using an innovative 2-phase fine-tuning depicted in Figure 3 – the models are first fine-tuned to a target language in general and then to the specific domain, in our case the oral history interviews.

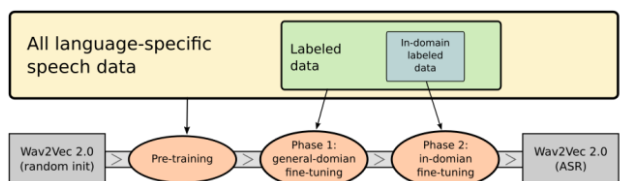

Figure 3: The scheme of 2-phase fine-tuning

The fine-tuning procedure was the same for all the languages, one of the key differences is in the choice of pre-trained model. For English, the model from Meta AI (`wav2vec2-base`) was used. For Czech and German, we have pre-trained our own model from scratch. The Czech model named CITRUS[7] was then

---

[5] https://archaeologydataservice.ac.uk/about/projects/atrium/

[6] https://lindat.mff.cuni.cz/services/uwebasr/
[7] https://huggingface.co/fav-kky/wav2vec2-base-cs-80k-CITRUS

re-used also as the base model for Slovak as this language is very similar to Czech.

Further details about the model training, including the description of the data used for fine-tuning, can be found in (Lehečka, 2023a) for English, Czech and German models and in (Lehečka, 2023b) for Slovak model.

Since the output from wav2vec models is a lower-cased, time-aligned continuous stream of words, further post-processing is needed in order to provide a high-quality human-readable text; this includes mainly case restoration, segmentation into sentences and adding of appropriate punctuation. All those post-processing steps are performed automatically, again using the latest NLP techniques employing the Transformer architecture – see (Švec, 2021) for details. The UWebASR service uses a simple HTTP API interface - the input data can be passed directly within the HTTP request or as a link to a file in the form of a URL. Live audio stream recognition from a given URL is supported as well. The output format includes plain text, machine-readable XML and JSON formats, and the WebVTT format for web captions. Recognition results (except TRS format) are streamed continuously.

While automatically generated transcripts and subtitles assist researchers in locating relevant interviews, they fall short in facilitating a comprehensive understanding of the entire testimony, whether through speech or text. Our innovative approach, again leveraging Transformer-based neural networks, seeks to bridge this gap. It not only aids in clearer navigation through lengthy testimonies but also transforms the listening experience from passive to interactive. By generating contextually relevant questions, our system enriches the interview monologues, allowing listeners to better orient themselves within the narrative and identify key segments of interest. These questions are designed to enhance understanding without altering the original meaning of the testimony, thereby maintaining the integrity of the historical record. Additionally, this method empowers users to engage more deeply with the material by posing their own inquiries, fostering a dynamic exploration of the rich narratives contained within the archives (Švec et al., 2024). This functionality is currently available outside of the LINDAT/CLARIAH-CZ portal in the test mode and will be integrated to the portal in the near future.

## 4. Do it yourself with Whisper

In Autumn 2022 OpenAI delivered Whisper[8] (Radford et al., 2022): an open source ASR toolkit that can handle about 100 different languages. OpenAI became famous with Chat GPT which was delivered a couple of months later and the enclosed LLM, but it also delivered a very good ASR engine that can be used in a (inter)national cloud environment (for example at a faculty), a local environment or at your local computer at home. The recognition velocity depends on the hardware used, but the recognition results are the same.

In the Netherlands, a first version of Whisper was tested by the end of 2022. We started to use it and made our colleagues in various ASR-projects enthusiastic. Whisper and its derivates, are basically a set of python instructions which can be installed on Windows, Linux, and Mac computers. Moreover, with a GPU in your computer, Whisper performs the recognition up to 10x faster.

OpenAI in the meantime, updated the environment and delivered in February 2023 a large model V2 (and some weeks later model V3) that increased the recognition results for most languages. Our impression is that V3 suffers more from hallucinations, which is confirmed by colleagues. For now, we therefore continue to use V2 as the "best" model.

The benefit of using Whisper is the open source character (more later) and the use of a powerful audio-conditional language model[9] (ACLM, Radford et al.).

When using Kaldi (Povey et al, 2011), we always had the problem that relatively unknown names or terms were not recognized. With Whisper, this is no longer an obstacle, because it uses an ACLM that "knows" these names and terms. Another advantage is that, unlike Kaldi, the recognized text includes punctuation, and capitalized words which greatly improves readability.

However, it should be noted that this applies to those languages that are "well" recognised, where the recordings are of good quality, people speak in a "standard" manner (no dialect) and contain no or little noise or background noise. If these conditions are less fulfilled, then, of course, the quality of transcriptions will also decrease. Yet Whisper surprises by often still offering a very usable result even in "worse" recordings.

Finally, the open-source character has the great advantage that not only OpenAI but also anyone with an eye for detail can use it to make it faster, better and richer. From February 2023, initiatives like WhisperX[10], Fast-Whisper[11], and others started working on improving Whisper's recognition procedure and making it more accessible.

For example, in February 2023, Jordi Bruin came up with MacWhisper[12], a very useful MacOS app using CPP conversion[13] of the original code. MacWhisper is an ideal tool for quickly creating a textual transcription (with or without timecode).

---

[8] https://openai.com/research/Whisper
[9] https://cdn.openai.com/papers/whisper.pdf
[10] https://github.com/m-bain/WhisperX
[11] https://github.com/SYSTRAN/faster-Whisper
[12] https://goodsnooze.gumroad.com/l/macWhisper
[13] https://github.com/ggerganov/Whisper.cpp

In October 2023 aTrain[14], a Windows-based version of Whisper, was delivered by the University of Gräz[15]. It is a fast and improved version of what was possible with Whisper (see Haberl et al., 2024).

Some "disadvantages" of Whisper are e.g. the not very accurate time estimation (when exactly was which word said) and the absence of speaker diarization. In the summer of 2023, Fast-Whisper already came with an even greater acceleration and in the autumn of 2023, WhisperX came with better time estimation and initial diarization. Since January 2024 diarization seems to be "a solved issue".

SURF in the Netherlands[16] had taken the initiative to establish Whisper as a service so that anyone with a SURF account could use it. Since a substantial number of researchers work with "sensitive" material, many research groups have set up a Whisper installation in their own secure network.

Google is working hard on updating an even better language model (Chirp, Universal Speech Model[17]) that will be able to handle 1000 different languages, and so did Meta with SeamlessM4T[18], but for now Whisper and its derivative versions, is the best candidate for a very good, fast and relatively easy ASR engine.

## 5. Remaining challenges

As remarked, transformer based speech models perform exceptionally better than the classical modular speech architectures such as Kaldi. This is especially true if a clean orthographic transcription with appropriate punctuation is the target of the speech to text conversion, which is the case for many research purposes. However, notably for linguistic research, more detailed output may be relevant. Which may require more than just this. Interviews are a relevant source e.g. for discourse analysis where pause durations and disfluencies play a paramount role in various methodological approaches See e.g. Van den Heuvel and Oostdijk, 2016). However, it are these phenomena that typically remain under the radar in standard ASR output (Lopez, Liesenfeld & Dingemanse, 2022). Disfluencies in spontaneous speech include *repetitions* (e.g. the the), *corrections* (e.g. Show me the flights … the early flights), *restarts* (e.g. There's a … Let's go), *filled pauses* (e.g. um and uh), and truncations resulting in *partial words* (e.g. wou- and oper-) (Lou & Johnson, 2020). The efforts in creating the transformed based large speech models are typically directed towards suppressing these phenomena in the text output they generate. Nonetheless, lately, there has been a growing interest in the advancement of technology that can decode speech while considering disfluencies. This technology is aimed at studying interruptions and corrections in communication settings. It has found various applications in the medical field, such as identifying early signs of cognitive decline (Claza et al., 2021, stuttering (Mitra et al, 2021), and detecting disfluencies through diagnostic tasks (Rohanian et al., 2021). Researchers are employing a combination of traditional AM/LM architectures and newer end-to-end models utilizing bidirectional LSTMs working in offline (non real-time) mode. These models incorporate features like word probabilities, confidence scores, prosodic features, pause duration statistics, and a range of acoustic features including fluency and speaking rate, which are now becoming standard in this area of research (Huang et al., 2018).

Another challenge is speaker diarisation which attributes the text output to resp. the interviewer and the interviewee(s). See for a review Park et al, (2022). Whisper-X is providing speaker diarisation in its output (Bain et al. 2023) as mentioned, but its performance needs to be explored and improved.

## 6. Conclusion

In this contribution we sketched a number of initiatives and toolkit approaches to improve automatic speech recognition for oral history interviews whilst offering these in a safe, well protected data shield minimizing dataleaks. We address the transcription portal and the webservices associated with speech processing at BAS, speech solutions developed at LINDAT, how to do it yourself with Whisper, but also mentioned a number of remaining challenges.

## 7. Bibliographical References

Baevski, A., Zhou, Y., Mohamed, A., & Auli, M. (2020). wav2vec 2.0: A framework for self-supervised learning of speech representations. *Advances in neural information processing systems*, 33, 12449-1246

Bain, M., Huh, J., Han, T., & Zisserman, A. (2023). WhisperX: Time-Accurate Speech Transcription of Long-Form Audio. ArXiv, abs/2303.00747.

Calzà, L., Gagliardi, G., Favretti, R. R., & Tamburini, F. (2021). Linguistic features and automatic classifiers for identifying mild cognitive impairment and dementia. *Computer Speech & Language*, 65, 101113.

Draxler, Chr. (2022) Automatic Transcription of Spoken Language Using Publicly Available Web Services. In: *FARE LINGUISTICA APPLICATA CON LE DIGITAL HUMANITIES, 2022*. https://bia.unibz.it/view/pdfCoverPage?instCode=39UBZ_INST&filePid=13284996250001241&download=true#page=28

Draxler, Chr., Van den Heuvel. H., Van Hessen, A., Calamai, S., Corti, L., Scagliola, S. (2020) A CLARIN Transcription Portal for Interview Data. In

---

[14] https://apps.microsoft.com/detail/9n15q44szns2?hl=en-US&gl=US

[15] https://doi.org/10.1016/j.jbef.2024.100891

[16] https://www.surf.nl/en

[17] https://cloud.google.com/speech-to-text/v2/docs/chirp-model

[18] https://ai.meta.com/blog/seamless-m4t/

*Proceedings of the 12th International Conference on Language Resources and Evaluation (LREC2020)*. pp. 3346-3352, http://www.lrec-conf.org/proceedings/lrec2020/pdf/2020.lrec-1.411.pdf

Draxler, Chr., Pömp, Julian (2022) OCTRA – An Innovative Approach to Orthographic Transcription. In *Proceedings INTERSPEECH 2022*, 5217-5218

Gref, M. (2022). *Robust Speech Recognition via Adaptation for German Oral History Interviews* (Doctoral dissertation, Universitäts-und Landesbibliothek Bonn). https://bonndoc.ulb.uni-bonn.de/xmlui/handle/20.500.11811/10373

Haberl, A., Fleiß, J., Kowald, D., & Thalmann, S. (2024). Take the aTrain. Introducing an interface for the Accessible Transcription of Interviews. *Journal of Behavioral and Experimental Finance*, 41, 100891. https://doi.org/10.1016/j.jbef.2024.100891

Huang, H. Y., Choi, E., & Yih, W. T. (2018). Flowqa: Grasping flow in history for conversational machine comprehension. arXiv preprint arXiv:1810.06683.

Kipp, A., Wesenick, M.-B., Schiel, F. (1997) Pronunciation Modeling Applied to Automatic Segmentation of Spontaneous Speech. In *Proceedings Eurospeech 1997*, 1023-1026, doi: 10.21437/Eurospeech.1997-358

Lehečka, J., Švec, J., Psutka, J.V., Ircing, P. (2023a) Transformer-based Speech Recognition Models for Oral History Archives in English, German, and Czech. In *Proceedings INTERSPEECH 2023*, 201-205, doi: 10.21437/Interspeech.2023-872

Lehečka, J., Psutka, J.V., Psutka, J. (2023b). Transfer Learning of Transformer-Based Speech Recognition Models from Czech to Slovak.. *TSD 2023. Lecture Notes in Computer Science 2023*, vol 14102, doi: 10.1007/978-3-031-40498-6_29

Lopez, A., Liesenfeld, A., & Dingemanse, M. (2022). Evaluation of Automatic Speech Recognition for Conversational Speech in Dutch, English and German: What Goes Missing? *In Proceedings of the 18th Conference on Natural Language Processing (KONVENS 2022)*. Potsdam.

Lou, P. J., & Johnson, M. (2020). End-to-end speech recognition and disfluency removal. arXiv preprint arXiv:2009.10298.

Mitra, V., Huang, Z., Lea, C., Tooley, L., Wu, S., Botten, D., Palekar, A., Thelapurath, S., Georgiou, P., Kajarekar, S., Bigham, J. (2021) Analysis and Tuning of a Voice Assistant System for Dysfluent Speech. In *Proceedings Interspeech 2021*, 4848-4852, doi: 10.21437/Interspeech.2021-2006.

Park, T. J., Kanda, N., Dimitriadis, D., Han, K. J., Watanabe, S., & Narayanan, S. (2022). A review of speaker diarization: Recent advances with deep learning. *Computer Speech & Language*, 72, 101317.

Povey, D., Ghoshal, A., Boulianne, G., Burget, L., Glembek, O., Goel, N., ... & Vesely, K. (2011). The Kaldi speech recognition toolkit. In *IEEE 2011 workshop on automatic speech recognition and understanding (No. CONF)*. IEEE Signal Processing Society.

Radford, A., Kim, J. W., Xu, T., Brockman, G., McLeavey, C., & Sutskever, I. (2022). Robust Speech Recognition via Large-Scale Weak Supervision (arXiv:2212.04356). arXiv. https://doi.org/10.48550/arXiv.2212.04356

Rohanian, M., Hough, J., & Purver, M. (2021). Alzheimer's dementia recognition using acoustic, lexical, disfluency and speech pause features robust to noisy inputs. arXiv preprint arXiv:2106.15684.

Švec, J., Lehečka, J., Šmídl, L., Ircing, P. (2021). Transformer-Based Automatic Punctuation Prediction and Word Casing Reconstruction of the ASR Output. TSD 2021. *Lecture Notes in Computer Science*, vol 12848, doi: 978-3-030-83527-9_7

Švec, J., Bulín, M., Frémund, A., Polák, F. (2024) Asking questions framework for oral history archives. *Accepted for ECIR 2024*

Scagliola, S., Corti, L., Calamai, S., Karrouche, N., Beeken, J., Van Hessen, A., Draxler, Chr., Van den Heuvel, H., Broekhuizen, M., Truong, K. (2020) Cross disciplinary overtures with interview data: Integrating digital practices and tools in the scholarly workflow. In: Simov, K., & Eskevich, M.: *Selected Papers from the CLARIN Annual Conference 2019* Leipzig, 30 September - 2 October 2019. Linköping Electronic Conference Proceedings 172:15, pp. 126-136. https://ep.liu.se/en/conference-article.aspx?series=ecp&issue=172&Article_No=15

Schmidt, Thomas & Wörner, Kai (2014) 'EXMARaLDA', in Jacques Durand, Ulrike Gut, and Gjert Kristoffersen (eds), *The Oxford Handbook of Corpus Phonology* (2014; online edn, Oxford Academic, 4 Aug. 2014), https://doi.org/10.1093/oxfordhb/9780199571932.013.030

Wittenburg, P., Brugman, H., Russel, A., Klassmann, A., & Sloetjes, H. (2006). ELAN: a professional framework for multimodality research. In *Proceedings of the 5th International Conference on Language Resources and Evaluation (LREC 2006)* (pp. 1556-1559), https://hdl.handle.net/11858/00-001M-0000-0013-1E7E-4

Van den Heuvel. H. & Oostdijk, N. (2016) Falling silent, lost for words ... Tracing personal involvement in interviews with Dutch war veterans. In *Proceedings of the 10th International Conference on Language Resources and Evaluation (LREC2016)*, Portorož, 23-28 May 2016. pp. 998-1001